# Unrealistic assumptions may lead to unrealistic simulation results: Droplet nuclei are neglected in a COVID-19 transmission simulation

# (Comments)

Masato Ida  (mida.0110@gmail.com)

Early retired from: *J-PARC Center, Japan Atomic Energy Agency*

Bale et al. [1] perform a numerical study of droplet/aerosol transport in the air to assess the probability of airborne transmission of COVID-19 from an infected person to a nearby healthy person. In their numerical study, the air flow field is solved by an implicit large eddy simulation model, and the airborne transport of the droplets/aerosols exhaled from an infected person is solved by a Lagrangian particle model which considers droplet/aerosol evaporation. In the model used for droplets/aerosols, I found that an unrealistic assumption is used which may have a significant impact on the numerical accuracy: Droplet nuclei contained in real respiratory droplets/aerosols are neglected.

As is known, respiratory droplets/aerosols exhaled from the human mouth and nose contain nonvolatile components such as salt, protein, and lipid [2, 3, 4, 5]. These components remain as ``droplet nuclei'' after the volatile components evaporate [3, 4, 6, 7, 8, 9, 10, 11, 12, 13, 14, 15]: See Ref. [5] for photos of evaporating respiratory droplets and residual droplet nuclei taken using an acoustic levitation technique. The size (or equilibrium diameter) of the droplet nuclei is not negligibly small compared to the initial diameter of the droplets/aerosols, being estimated to be about 0.2~0.4 times the initial diameter at ~50% relative humidity [16, 5]. The droplet nuclei are often considered to be the virus transmitter in airborne transmission of infectious diseases.

On the contrary, the model used by Bale et al. does not consider the droplet nuclei. Because the right-hand side of the model equation for the droplet mass (the lower equation in 17: See Ref. [17] for detailed review of this type of models) is negative for evaporation, the masses of the droplets/aerosols decrease without bound towards zero. This is due to the assumption that no, or negligibly small amount of, nonvolatile components are contained in the droplets/aerosols. Figure 4 of Bale et al.'s paper clearly demonstrates that the droplet diameter decreases without bound when the model equation in 17 is used.

This unrealistic assumption may lead to significant errors in the numerical simulations. First of all, the size of the droplets/aerosols is significantly underestimated particularly when the simulation time (up to 60 min in Bale et al.'s study) is much longer than the evaporation time of the droplets/aerosols (< 1 s for 10 μm droplets and about 10 s for 100 μm droplets at 293 K ambient temperature and 50% relative humidity [18]). After a droplet with nonvolatile components evaporates and becomes a droplet nucleus, its size is fixed. If the nonvolatile components are neglected, however, the size of the droplet decreases without bound by continuous evaporation.

This underestimation of droplet/aerosol size would have a strong impact on Bale et al.'s numerical results. As is well known in many research fields including fluid dynamics and infectious diseases, the transport of droplets/aerosols in the air and in the human body is strongly dependent on the droplet/aerosol size among others [3, 4, 6, 7, 10, 11, 12, 14, 19]. Inaccurate estimation of droplet/aerosol size could thus result in a large error.

Since in the COVID-19 pandemic there has been much debate on the characteristics of respiratory droplets/aerosols [4, 5, 6, 7, 8, 14, 15, 18, 19], careful discussion is necessary. Bale et al. should clarify:

(a) why they used the model that does not consider the droplet nuclei even though more realistic models that consider the droplet nuclei have been used in previous works by other researchers [3, 12, 4], and
(b) how their numerical results (e.g., the transport dynamics of droplets/aerosols and the

resulting infection probability) change if the droplet nuclei are taken into account.

In addition, several important variables (e.g., $Nu$, $Pr$, and $Sc$ in Eq. 17) are not explained in Bale et al.'s paper. The authors must be more careful in writing a paper.

**Additional Information**

The author declares no competing interests.